\documentclass[aps, prl, reprint, groupedaddress, superscriptaddress, footinbib, floatfix]{revtex4-1}

\usepackage{amsmath, amssymb, bm}
\usepackage{graphicx}
\usepackage{natbib}
\usepackage{soul}
\usepackage{float}
\usepackage{upgreek}
\usepackage[euler]{textgreek}
\usepackage[english]{babel} 
\usepackage{blindtext}
\usepackage[dvipsnames]{xcolor}
\usepackage{braket}
\usepackage{hyperref}
\definecolor{ddblue}{RGB}{0,0,160}
\hypersetup{colorlinks=true,linktoc=all,linkcolor=ddblue,citecolor=blue,urlcolor=[rgb]{0, 0.38, 0}}

\newcommand*\vecc[1]{\ifmmode\bm{#1}\else\textbf{#1}\fi}

\begin{document}
	
	\preprint{APS/123-QED}
	
	\title{Collectively enhanced Ramsey readout by cavity sub- to superradiant transition}

	\author{Eliot Bohr}
	\affiliation{Niels Bohr Institute, University of Copenhagen, DK-2100 Copenhagen, Denmark}
	\author{Sofus L. Kristensen}
	\affiliation{Niels Bohr Institute, University of Copenhagen, DK-2100 Copenhagen, Denmark}
        \author{Christoph Hotter}
	\affiliation{Institut f\"ur Theoretische Physik, Universit\"at Innsbruck, Technikerstr. 21a, A-6020 Innsbruck, Austria}
	\author{Stefan Alaric Sch\"affer}
	\affiliation{Niels Bohr Institute, University of Copenhagen, DK-2100 Copenhagen, Denmark}
        \author{Julian Robinson-Tait}
	\affiliation{Niels Bohr Institute, University of Copenhagen, DK-2100 Copenhagen, Denmark}
    \author{Jan W. Thomsen}
	\affiliation{Niels Bohr Institute, University of Copenhagen, DK-2100 Copenhagen, Denmark}
    \author{Tanya Zelevinsky}
	\affiliation{Department of Physics, Columbia University, 538 West 120th Street, New York, NY 10027-5255, USA}
	
	\author{Helmut Ritsch}
	\affiliation{Institut f\"ur Theoretische Physik, Universit\"at Innsbruck, Technikerstr. 21a, A-6020 Innsbruck, Austria}
	\author{J\"org Helge M\"uller}
	\affiliation{Niels Bohr Institute, University of Copenhagen, DK-2100 Copenhagen, Denmark}

	\date{\today}
	
	\begin{abstract}
		
            When an inverted ensemble of atoms is tightly packed on the scale of its emission wavelength or when the atoms are collectively strongly coupled to a single cavity mode, their dipoles will align and decay rapidly via a superradiant burst.  However, a spread-out dipole phase distribution theory predicts a required minimum threshold of atomic excitation for superradiance to occur. Here we experimentally confirm this predicted threshold for superradiant emission on a narrow optical transition when exciting the atoms transversely and show how to take advantage of the resulting sub- to superradiant transition.  A $\pi/2$-pulse places the atoms in a subradiant state, protected from collective cavity decay, which we exploit during the free evolution period in a corresponding Ramsey pulse sequence. The final excited state population is read out via superradiant emission from the inverted atomic ensemble after a second $\pi/2$-pulse, and with minimal heating this allows for multiple Ramsey sequences within one experimental cycle. Our scheme is a fundamentally new approach to atomic state readout characterized by its speed, simplicity, and high sensitivity. It demonstrates the potential of sensors using collective effects in cavity-coupled quantum emitters.
 
	\end{abstract}

	\maketitle
 
 The precise measurement and manipulation of atomic states lie at the heart of a broad range of scientific and technological advancements. Quantum sensors such as atomic clocks, with their unprecedented accuracy and stability \cite{campbell2017,brewer2019, McGrew2018,Bloom2014,Ushijima2015,Peik2016}, have revolutionized fields ranging from fundamental physics to global positioning systems. The development of novel state detection methods plays a pivotal role in improving the performance and capabilities of these atomic sensors \cite{Vallet2017,Chen2014,Lodewyck2009, Orenes2022, Bowden2020, Corgier2023}. In recent years, quantum technologies have emerged as a promising avenue for enhancing sensitivities in measurement and sensing applications. Among these technologies, ultracold atomic systems, combined with high-finesse optical cavities, have demonstrated remarkable potential due to their ability to access and manipulate collective quantum states \cite{bohnet2012steady, Norcia2016,Norcia16mill2, Hemmerich2019, Schaffer2022, Kristensen2023, Leroux2010, Colombo2022}.

In this Letter, we present the experimental realization of a novel state detection method \cite{hotter2023} that exploits the phenomenon of superradiant (SR) light emission \cite{dicke1954, gross1982superradiance} after transverse Ramsey interrogation of atoms in an optical cavity. We experimentally demonstrate the excitation threshold for superradiant states and employ the interplay of sub- and superradiance for a fast cavity-assisted readout. Ramsey spectroscopy \cite{ramsey1950molecular}, widely used for precision measurements \cite{Ludlow2015,muller2018} in atomic systems, involves a two-pulse sequence that measures the accumulated phase difference between a laser and a superposition of atomic states. By incorporating an optical cavity into the experimental setup, we harness the collective nature of randomly distributed atoms to generate a state-selective superradiant decay into a well-defined cavity mode. This drastically increases detection efficiency compared to free-space spontaneous emission detection, while not requiring any additional lasers for electron shelving. We demonstrate the reusability of our atomic ensemble by turning on cooling light in between successive Ramsey sequences and show 100s of repetitions on the same atomic cloud, ultimately limited by background gas collisions of the vacuum chamber. Our experimental system comprises ultracold strontium atoms trapped within a high-finesse optical cavity, in the so-called bad-cavity regime. The coherent excitation and de-excitation of the atomic ensemble, induced by carefully timed optical pulses, generate an enhanced superradiant emission into the cavity mode when the inversion exceeds threshold, or a subradiant emission \cite{Kaiser2016,Glicenstein22,Albrecht_2019,Tiranov2023,Orioli2022,Ansejo2017,zanner2022coherent,holzinger2022control} when below. This collective emission serves as a sensitive probe for detecting and characterizing the atomic states, enabling improved state discrimination and measurement precision. In addition, the subradiant behavior allows for long interaction times necessary for resolving and exploiting ultranarrow clock transitions.

The use of ultracold strontium atoms in conjunction with an optical cavity presents several advantages for our proposed state detection method. The long coherence times and precise control achievable in ultracold atomic systems facilitate the generation of highly coherent population oscillations. The optical cavity provides enhanced light-matter interaction, amplifying the superradiant emission and enabling efficient state readout. 

\begin{figure}[t!]
\centering
\includegraphics[width=1\linewidth]{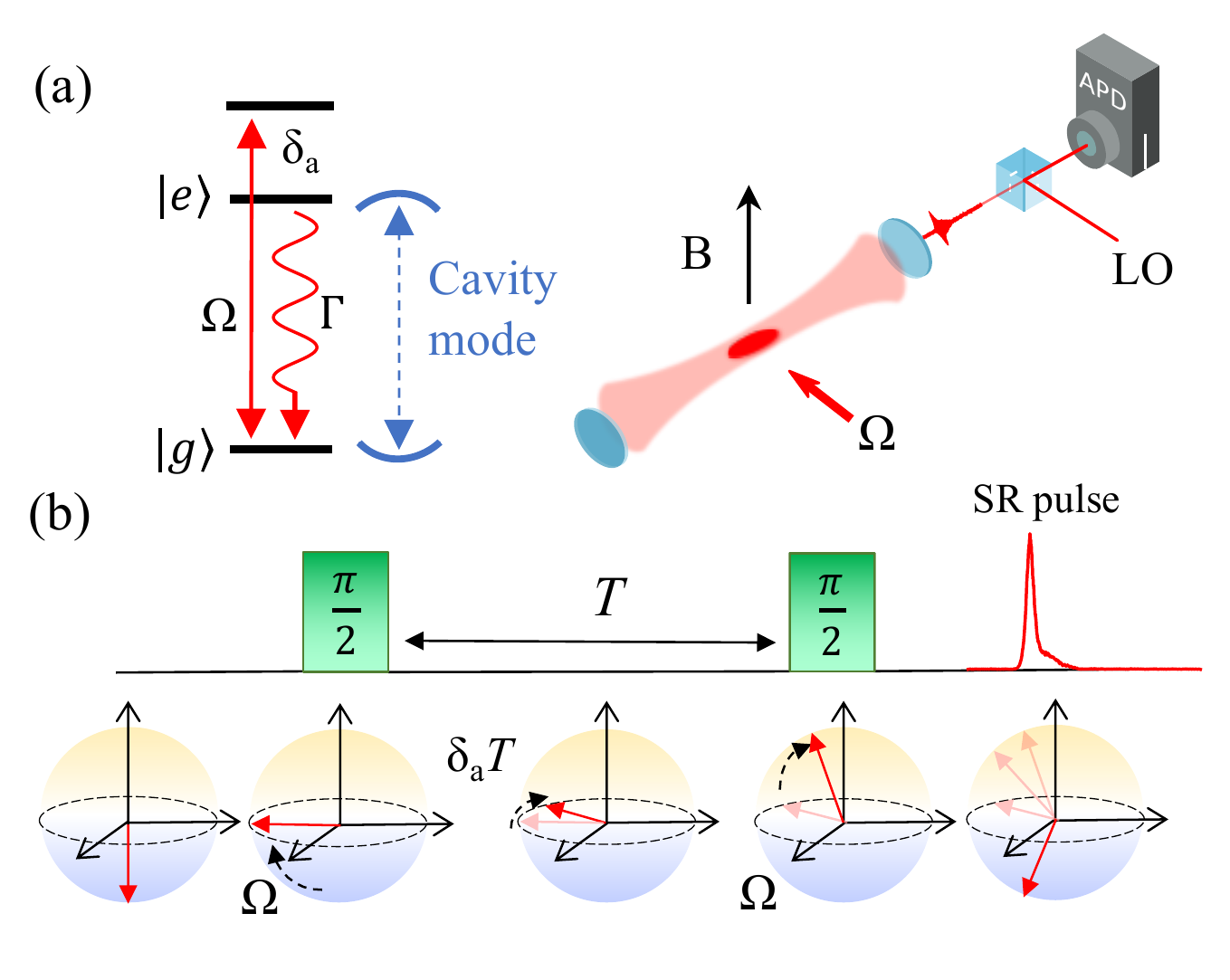} 
\caption{(a) Relevant level diagram and experimental schematic. We cool and trap strontium atoms in the center of the cavity. We then pump the atoms at a laser detuning, $\delta_{a}$, from the $^1$S$_0$ $\rightarrow$ $^3$P$_1$, $m_{j}=0$ transition transversely to the cavity axis and observe the intensity of the emitted pulse via a beat detection with a local oscillator. (b)  Diagram of a Ramsey interrogation sequence which uses two $\pi/2$-pulses separated by a free evolution time $T$.  Bloch spheres below the sequence show the collective Bloch vector at corresponding times in the sequence which can result in a SR pulse.  The first $\pi/2$-pulse pulse brings the atoms to a 50\% fractional excitation, at which point the excitation is protected from cavity emission during the free evolution time.  A second $\pi/2$-pulse brings the excitation above or below 50\% depending on the phase accumulated, $\delta_aT$.  If positive inversion is reached, a SR pulse is emitted, shown as a collective Bloch vector accelerated downward on the final Bloch sphere, with peak intensity and photon number corresponding to the amount of positive inversion.}
\label{fig1}
\end{figure}

\emph{Experimental setup.}\textemdash We cool a cloud of up to $N = 4 \times 10^7$ $^{88}$Sr atoms down to 2 $\mu$K using a two-stage magneto-optical trap (MOT). The cloud of atoms is then centered in the fundamental mode of an optical cavity (Fig. \ref{fig1}) with finesse $F\approx 1000$ and cavity decay rate $\kappa/2\pi = 780$ kHz. By locking the cavity to a reference laser one free spectral range away, we tune a TEM$_{00}$ resonance of the cavity to the $\ket{e}$ = $^3$P$_1$, $m_{j}=0$ $\rightarrow$ $\ket{g}$ = $^1$S$_0$ transition frequency, $\omega_a$. The cavity mode waist radius is $450$~$\mu$m, and the atomic cloud has a vertical full height of 100 $\mu$m and horizontal full width of 200 $\mu$m, fitting well within the TEM$_{00}$ cavity mode volume. 

The atoms are randomly distributed across the cavity nodes and antinodes, resulting in inhomogeneous single-atom couplings, $g$. At positions of maximal coupling to the cavity, the system is in the regime of small single-atom cooperativity with Purcell factor $C = 4.4 \times 10^{-4} \ll 1$ but large collective cooperativity $NC \gg 1$.  A highly-resolved normal-mode splitting confirms that the collective vacuum Rabi frequency exceeds the relevant decoherence rates in the system: $2g\sqrt{N} \gg \kappa,\gamma$.  Averaging the coupling over the standing wave mode of the cavity reduces the collective cooperativity by a factor of 2. Taking into account the finite size of the atomic cloud compared to the cavity waist leads to a further slight reduction of the effective coupling \cite{Hu2015}.

After cooling the atoms, we turn off the MOT laser beams and quadrupole magnetic field, leaving only a vertical bias field from a set of Helmholtz coils.  This field of 0.2 mT provides a quantization axis and separates each of the $m_j=\pm1$ Zeeman sublevels of $^3P_1$ by a splitting of $\Delta = 2\pi \times 4.2$ MHz.  This together with the choice of polarization direction along the bias field, reduces the dynamics to a simple two-level system. The atoms are driven transversely to the cavity axis, as illustrated in Fig. \ref{fig1}a, with a laser at frequency $\omega_l$, detuning $\delta_a=\omega_l-\omega_a$, and corresponding Rabi frequency $\Omega = 2\pi \times (833 \pm 30)$ kHz. We assume a uniform Rabi frequency since the pump laser waist is much larger than the atomic cloud and it is large compared to the Doppler width of the cold sample. Emitted power into the cavity mode is detected on an avalanche photodetector via a heterodyne beat measurement with a stable local oscillator (LO) to reject the cavity locking signal.

\begin{figure}[t!]
\centering
\includegraphics[width=1\linewidth]{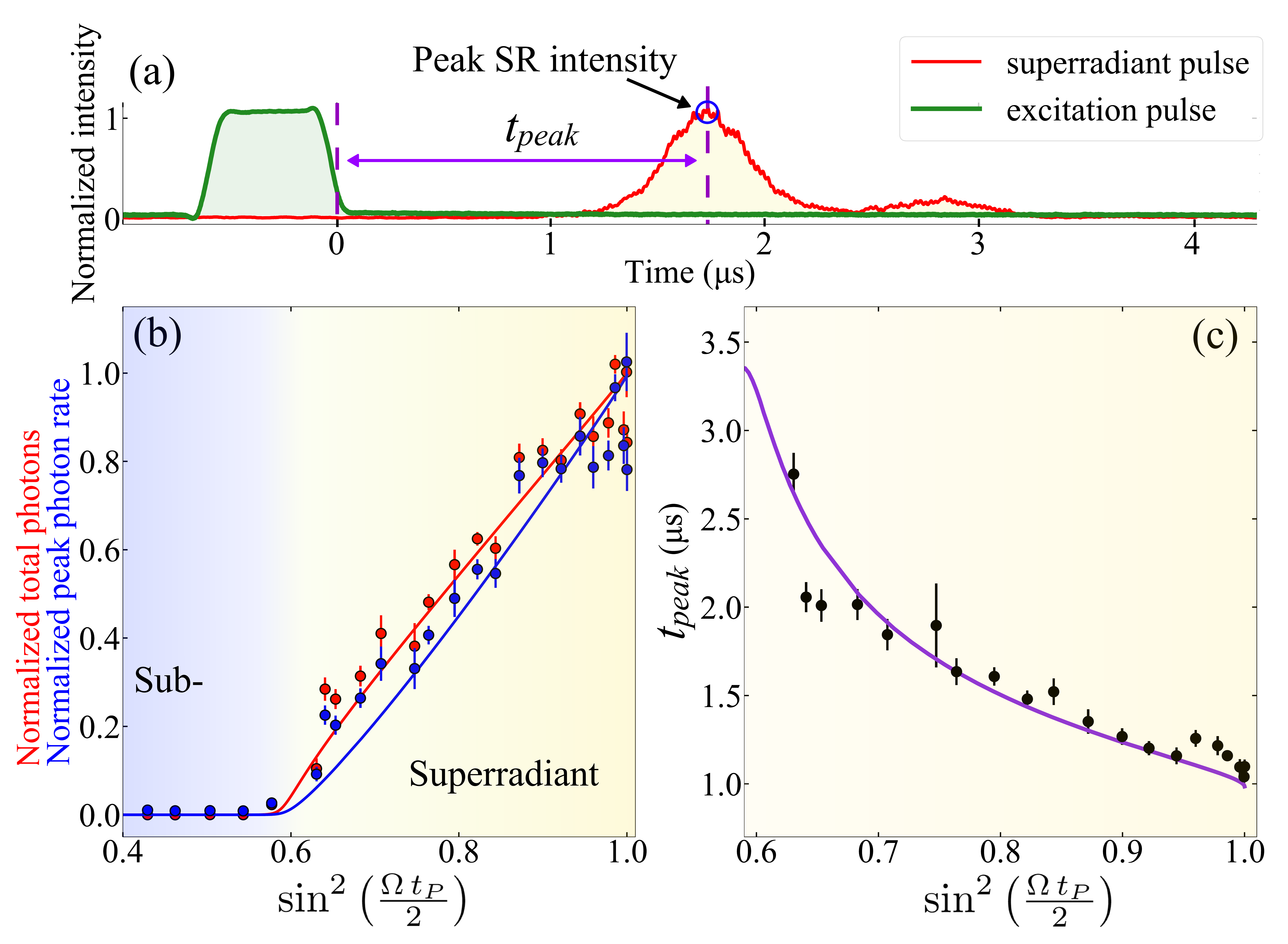} 
\caption{(a) A single-shot trace showing an excitation pulse (green), pulse delay time (purple), and detected SR light pulse (red) along the cavity mode with peak emitted intensity circled in blue. (b) Normalized emitted pulse area (red) and peak amplitude (blue) for varying excitation pulse duration.  For excitation pulse angles $\Omega \, t_{P} \leq 0.57 \pi$ no SR pulses are observed, while above this threshold the SR peak amplitude and area scale linearly with pulse angle.  Solid lines are simulations based on the experimental parameters. (c) Delay times of the SR emission for varying pulse angle. The solid purple line is a simulation treating $g$ as a free parameter. Each data point in (b) and (c) is an average of ten measurements with error bars represented by the standard deviation of the mean.}
\label{fig2}
\end{figure}

\emph{Threshold results.}\textemdash We investigate SR emission after applying a single pulse of resonant light with varying time duration, $t_{P}$, corresponding to different excitation angles on the Bloch sphere (see Fig. \ref{fig1}b).  For this investigation, we reduced the atom number to $N = 2 \times 10^7$ to reduce the collective Rabi frequency so that SR emission does not start during the excitation pulse.   Figure \ref{fig2}a shows a single-shot trace of SR emission after a nominal $3\pi/4$-pulse with peak emitted intensity circled in blue.  In Fig. \ref{fig2}b, we plot the normalized average peak emitted intensity (blue) and integral of the pulse (red) of ten traces for various excitation angles on the Bloch sphere. On the x-axis, we plot $\sin^{2}{\left(\frac{\Omega \, t_{P}}{2}\right)}$, as a proxy for the excited state population at the end of the pump pulse, $\langle \sigma^{22} \rangle_{t=0}$. The maximum population inversion we can achieve is limited by the decoherence in the system (spontaneous emission and Doppler dephasing) during the pump pulse duration.  We experimentally determine a nominal $\pi$-pulse as the pump pulse duration which results in the largest peak SR emission. 

We observe a threshold excitation angle for a SR burst to be emitted into the cavity mode. Above this threshold, both the peak intensity and pulse area increase approximately linearly, in agreement with the simulation as shown by the solid lines for the experimental parameters. The linear behavior of the peak intensity is attributed to the large collective interaction rate overcoming the cavity decay rate, $2g\sqrt{N} > \kappa$. As a result, the excitations oscillate back and forth between the atoms and cavity before leaving the cavity, as manifest in the slight revival of cavity emission visible in Fig. \ref{fig2}a. In this oscillatory SR regime, the peak scales linearly rather than quadratically with the excited state atom number \cite{Schaffer2022}. Due to imperfections in the $\pi$-pulse excitation, measured peak powers near the maximum tend to be biased towards lower values. Slight deviations from a $\pi$-pulse always result in lower peak emitted intensity. As a result, measurements up to approximately $\sin^{2}{\left(\frac{\Omega \, t_{P}}{2}\right)} = 0.9$ exhibit a high degree of agreement, while the agreement gradually levels off beyond this value.

In a system without dephasing, one would expect a threshold at 
$\Omega \, t_{P} = 0.5\pi$, corresponding to $\langle \sigma^{22} \rangle_{t=0} = 50\%$, but due to decoherence in the time before the pulse is emitted, the threshold for superradiance is found to be $\Omega \, t_{P} = 0.57 \pi$ both in experiment and in simulation, as shown in Fig. \ref{fig2}b. Immediately after the SR pulse is emitted, the atoms retain an excited state population of $1-\langle \sigma^{22} \rangle_{t=0}$, leaving them in a subradiant state. If the initial excited state population is below threshold, the atoms are not able to synchronize and their emission into the cavity is suppressed \cite{hotter2023}.  

Figure \ref{fig2}c shows the delay time between the end of the pump pulse and the peak intensity of the emitted pulse. This is the time required for the individual atomic dipoles to synchronize through the cavity field according to each atom's position in the cavity mode.  In agreement with previous research \cite{Hemmerich2019}, larger inversions result in faster emissions.  The purple solid line corresponds to a simulation that treats $g$ as a free parameter, and aligns well with our experimental data for $g = 2\pi \times 450$ Hz. A normal-mode splitting measurement suggests a value of $g = 2\pi \times 635$ Hz. The disagreement is likely due to simplifying assumptions in the model.

If the atoms are excited through the cavity mode, the phases of the atomic dipoles will by necessity match their location in the cavity mode coupling so as to constructively interfere with the cavity mode. In this case, there is no threshold for superradiance and the delay time will be much shorter.

\begin{figure}[t!]
\centering
\includegraphics[width=1\linewidth]{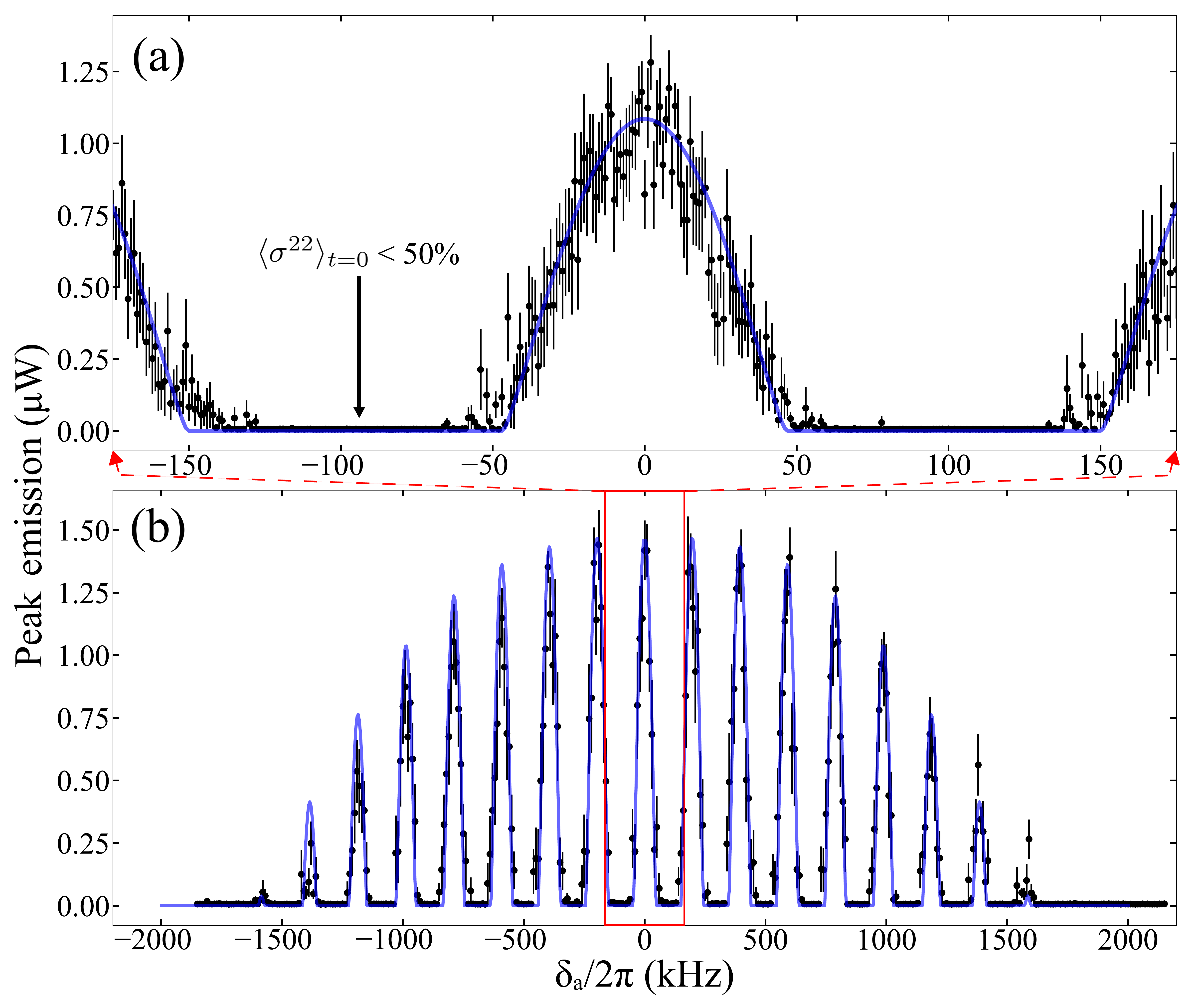} 
\caption{We excite atoms with varying laser detunings and detect SR peak emissions along the cavity axis. (a) A narrow scan around the center fringe, boxed in red in (b), taken with high resolution. (b) Scan taken over the complete spectroscopic lineshape. The blue solid lines are fits from simulation with an overall rescaled amplitude. Each data point is a mean of ten measurements, each with a separate MOT loading cycle.  Error bars indicate standard deviations of the mean.}
\label{fig3}
\end{figure}

\emph{Ramsey lineshape results}.\textemdash We can exploit the feature that a transverse $\pi/2$-pulse excitation places the sample in a subradiant state, protected from collective decay into the cavity mode. 
When combined with a SR pulse readout after a second $\pi/2$-pulse, this allows for a cavity-assisted Ramsey scheme. By varying the pump laser detuning, $\delta_{a}$, and detecting the power emitted from the cavity mode, we plot the average peak amplitude and map out a spectroscopic fringe pattern signal shown in Fig. \ref{fig3}.  Each data point is an average of ten repeated measurements, and each measurement is from a separate MOT loading cycle.  
Figure \ref{fig3}a is a scan around the central Ramsey fringe taken with higher spectral resolution while Fig. \ref{fig3}b is data taken over the complete lineshape.

The blue solid lines are simulations based on the experimental parameters with a rescaled amplitude. The data show strong agreement with the lineshape predicted by simulation.  The width of the envelope of the fringe pattern is given by $\Delta f = 1/\tau_{P} = 3.33$ MHz, where $\tau_{P} = 300$ ns is the duration of each $\pi/2$-pulse. We select a free evolution time $T= 5$ $\mu$s to balance between being significantly shorter than the natural decay time of $22$ $\mu$s required for a sizable signal, but long enough to observe numerous fringes. The frequency spacing of the Ramsey fringes, or free Ramsey range (FRR), is given by the inverse of the interpulsar free evolution time, $1/T = 200$ kHz.  Different from the traditional Ramsey lineshape, the superradiance enhanced lineshape is truncated for detunings which do not result in positive atomic population inversion.  This yields flat near zero-photon emission zones in between the fringes. In Fig. \ref{fig3}b, we note a slight asymmetry in the data which can be attributed to residual detuning-dependent pumping intensity.

Rabi-type interrogations with durations comparable to the decay rate are limited by SR emission, which begins as soon as positive inversion is achieved. The subradiant behavior of the atoms is essential and ensures protection against cavity decay until after the final $\pi/2$-pulse is applied.  For Ramsey excitation through the cavity, the protection via subradiant states is lacking due to the inherent phase-matching, significantly reducing the possible free evolution time.  

In a locking scheme, the laser frequency or phase can be incrementally adjusted around a fringe to increase the sensitivity to detuning. By analyzing consecutive measurements, it is possible to generate a feedback signal that guides the laser frequency toward the atomic resonance. Notably, the collectively enhanced lineshape exhibits a distinct kink at the sub- to superradiance transitions, which may serve as a highly precise and narrow feature for laser locking purposes.  Currently, the observed statistics of the peak amplitude around these kinks show less signal-to-noise than around the center of the fringes.

\emph{Nondestructive readout results.}\textemdash Each Ramsey sequence causes maximally two recoils per atom, one from the two Ramsey excitation pulses (perpendicular to the cavity axis), and one from the SR emission (along the cavity axis).  Without cooling in between sequences, we can conduct about ten sequences on the same atomic ensemble.  However, each measurement has reduced contrast due to heating, and the resonance shifts each sequence as the atoms are accelerated away from the pump laser direction.

\begin{figure}[t!]
\centering
\includegraphics[width=1\linewidth]{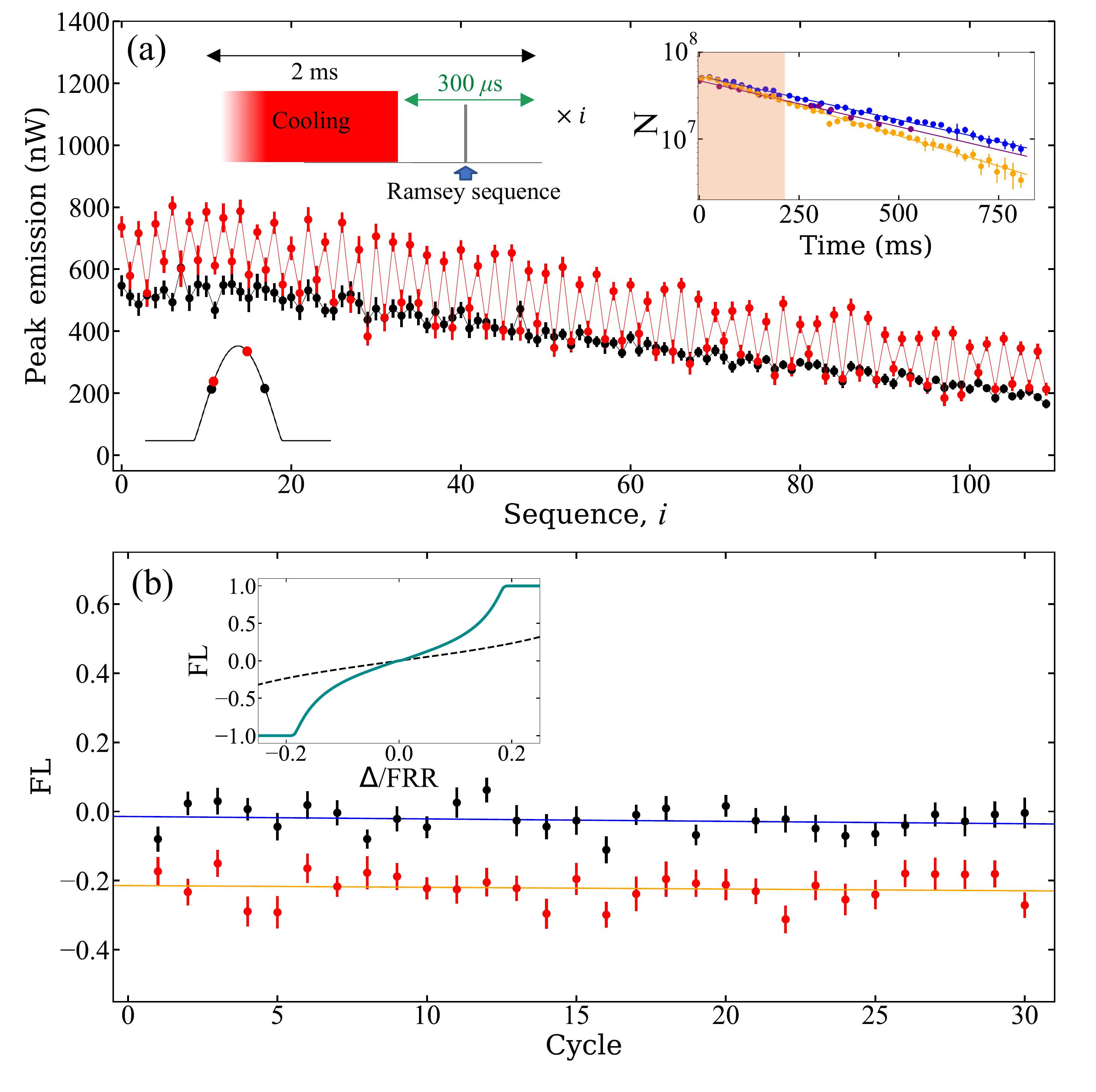} 
\caption{(a) Peak SR emitted intensities while stepping the interrogation laser frequency between Ramsey sequences in the case of a resonant (black) and detuned laser (red) with error bars representing the standard deviation of the mean of 30 MOT loading cycles. Top left inset: timing diagram for interleaved cooling and Ramsey interrogation.  We can perform more than 100 repetitions within a MOT loading cycle. Top right inset: atom number measurements via shadow imaging taken at different times after the start of the repeated Ramsey measurements for the case of blocked Ramsey pump pulses (blue), on-resonant Ramsey pulses (yellow), and for reference, leaving on the MOT the entire time with no pump pulses (purple).  Each data point is an average of ten measured atom numbers with error bars representing the standard deviation of the mean. (b) Frequency locator averaged over the 108 pulses in each trace taken in pairs. Error bars are standard deviations of the mean of the 54 pairs of pulses.  Solid lines are best-fit lines, consistent with zero slopes, proving no frequency chirp or systematic detuning for 30 loading cycles taken at 6 s intervals. Inset: conversion of FL to laser detuning given our step size and fringe width for the collectively enhanced Ramsey lineshape (cyan) and a traditional Ramsey lineshape (black dotted) for reference. }
\label{fig4}
\end{figure}

To remove the systematic Doppler shift resulting from consecutive Ramsey sequences,  we employ a 2 ms cycling time comprising 1.7 ms of cooling on the single-frequency 689 nm MOT, and 300 $\mu$s for Ramsey interrogation and readout. The timing is illustrated in the top left inset in Fig. \ref{fig4}a. The top right inset shows the atom number estimated from shadow images after a varying number of 2 ms sequences, with Ramsey excitation pulses (yellow) and without (blue), as well as with the 689 nm MOT left on continuously (purple).  We observe a decay constant of 313 $\pm$ 4 ms when resonant excitation pulses are applied, and 431 $\pm$ 9 ms when the excitation pulses are blocked.  This indicates a slight reduction in recapture due to the heating from the resonant Ramsey sequences. The decay constant of the atoms with the 689 nm MOT left on continuously is 409 $\pm$ 9 ms, limited by background gas collisions.  Therefore, turning off the MOT for 300 $\mu$s during each sequence does not result in increased losses. Also, it is possible to get 100s of useful pulses with millions of atoms within a single MOT loading cycle.

A 2 ms cooling and interrogation sequence is much shorter than the $\sim$ 1 s time to trap and cool a completely new atomic ensemble, drastically increasing the repetition rate. The Ramsey sequence time is still short compared to the 2 ms, and is limited by the natural lifetime of $^3$P$_1$. However, this interrogation time can be longer than the cooling time if we use a more narrow clock transition such as $^3$P$_0$ and place the atoms in a lattice to counteract gravity.

To demonstrate a preliminary clock feedback measurement, we step the frequency of the interrogation laser between Ramsey sequences within a MOT loading cycle.  In Fig. \ref{fig4}a we show SR peak intensities as we toggle the frequency symmetrically around the resonance (black) and around a detuned frequency (red). Averaging over 30 MOT loading cycles, it is evident that in the off-resonant case, the frequency step nearer to atomic resonance consistently yields higher peak emissions.  Both sets exhibit a decrease in intensity due to the gradual loss of atoms over time.

We define a frequency locator, FL = $\langle \frac{P_{i+1}-P_{i}}{P_{i+1}+P_{i}} \rangle$, where $P_{i}$ is the peak intensity of the $i$th pulse within a cycle.  The FL is the difference between consecutive peaks divided by the sum, to normalize for variation in atom number. We average this value in pairs of consecutive Ramsey sequences over an experimental cycle.  Given the free evolution time and frequency stepping size, this value can be converted to an interrogation laser detuning and used for laser frequency feedback. The inset shows how the FL can be related to a detuning when the step size is $0.1$~FRR, where the solid line corresponds to the collectively enhanced lineshape and the less steep dotted line is derived from a traditional Ramsey lineshape for reference.  In Fig. \ref{fig4}b, we plot the FL averaged over 108 pulses for each MOT loading cycle.  Best-fit lines show slopes consistent with zero for both sets, indicating there is no systematic change in frequency throughout the 30 MOT loading cycles taken at 6 s intervals.  The off-resonant case has a constant offset, signifying a constant interrogation laser detuning. The 30 traces are taken over the course of three minutes and the deviations in the data points represent frequency excursions of the interrogation laser on this time scale. 
 
The observed readout noise is above the atomic projection noise. How much of this excess noise is due to interrogation laser phase fluctuations and how much stems from inherent stochastic dynamics of SR emission is an important question for applications and is subject to further investigation.

\emph{Concluding remarks.}\textemdash  Our study demonstrates the use of collective atomic decay of ultracold Sr for state detection. The collective atomic interaction behaves in agreement with a sub- to superradiant phase transition with respect to emission in the cavity mode as a function of the ensemble inversion.  This behavior can offer a superradiant enhanced readout of a Ramsey interrogation sequence, producing a fast and directional state readout with no additional lasers. This makes the scheme useful for a wide range of sensors.

If we apply the technique to an ultranarrow clock transition, such as the $^3$P$_0$ transition in Sr or Yb, the time spent measuring the atomic state could exceed the necessary cooling times, which could greatly reduce the Dick effect \cite{Dick1987}. To implement this approach in optical lattice clocks, we propose adding a separate state-readout cavity perpendicular to the necessary 1D lattice confinement axis. One could also implement the method in a shelving scheme where atoms are first shelved on the $^3$P$_0$ state, before a superradiant readout on the $^3$P$_1$ transition.

Further investigations will focus on reducing shot-to-shot fluctuations by quantifying the competing emission modes that contribute to the inherent pulse statistics of the SR signal. By combining characteristics such as delay time and peak intensity, a more precise determination of atomic inversion may be obtained. The immediate benefits offered by our scheme could be applied to any quantum sensor relying on particularly the population difference readout of a quantum state. By judicious choice of parameters such as finesse and cooperativity of the readout cavity, our scheme can be implemented in other types of sensors using quantum emitters of varying ensemble sizes. 

We thank Mikkel Tang and Asbjørn A. Jørgensen for contributions to the experimental apparatus as well as helpful discussions. This project was supported by the European Union’s (EU) Horizon 2020 research and innovation program under the Marie Sklodowska-Curie Grant Agreement No. 860579 (MoSaiQC) and Grant Agreement No. 820404 (iqClock project), the USOQS project (17FUN03) under the EMPIR initiative, and the Q-Clocks project under the European Commission’s QuantERA initiative. We acknowledge funding from VILLUM FONDEN via Research Grant No. 17558. 

E. B. and S. L. K. contributed equally to this work.

	\bibliographystyle{apsrev4-1}
	\bibliography{biblio_cavity_ramsey}
	

\end{document}